\documentclass[a4paper]{jpconf}
\usepackage{graphicx}
\begin{document}
\title{Review of the Laguerre-Gauss mode technology research program at Birmingham}

\author{P. Fulda, 
C. Bond, 
D. Brown,
F. Br\"{u}ckner
L. Carbone, 
S. Chelkowski\footnote{Presently at: Jenoptik AG, Carl-Zeiss-Strasse 1, 07739, Jena, Germany}, 
S. Hild\footnote{Presently at: Institute for Gravitational Research, University of Glasgow, G12 8QQ, UK}, 
K. Kokeyama\footnote{Presently at: Physics and Astronomy, Louisiana State University, Baton Rouge, LA 70803, USA}, \\
M. Wang,
and A. Freise}

\address{School of Physics and Astronomy, University of Birmingham\\
Edgbaston, Birmingham, B15~2TT, UK}
\ead{pfulda@star.sr.bham.ac.uk}

\begin{abstract}
Gravitational wave detectors from the advanced generation onwards are expected to be limited in sensitivity by thermal noise of the optics, making the reduction of this noise a key factor in the success of such detectors. 
A proposed method for reducing the impact of this noise is to use higher-order Laguerre-Gauss (LG) modes for the readout beam, as opposed to the currently used fundamental mode. 
We present here a synopsis of the research program undertaken by the University of Birmingham into the suitability of LG mode technology for future gravitational wave detectors. 
This will cover our previous and current work on this topic, from initial simulations and table-top LG mode experiments up to implementation in a prototype scale suspended cavity and high-power laser bench. 
\end{abstract}

\section{Introduction}
Advanced gravitational wave detectors such as Advanced LIGO \cite{Harry10} and Advanced Virgo \cite{VIR-027A-09} are being constructed with the goal of a tenfold increase in sensitivity over the first generation of detectors \cite{Sigg04}\cite{virgo2006short}. This is expected to be sufficient to make the first conclusive gravitational wave detection.
The projected sensitivity of both Advanced LIGO and Advanced Virgo will be limited over the observation frequency band by several fundamental noise sources: seismic noise, radiation pressure noise, suspension thermal noise, shot noise, and thermal noise of the test masses themselves \cite{Rowan05}. 
The dominant contribution to the total thermal noise of the test masses is expected to be from the Brownian motion of the atoms that make up the reflective coatings. This motion causes an uncertainty in the phase of the light reflected from the test mass, and will therefore appear as noise in the detector output. 
This thermal noise has until now not been a limiting noise source in detectors, but as progress within the field pushes other noise sources ever lower, its mitigation becomes a more urgent priority.

One way to reduce the effects of the coating Brownian thermal noise is to use a larger beam spot on the test mass surfaces, since a larger beam averages more effectively over the random motions of the surface. 
The coating Brownian thermal noise power spectral density has been shown to scale with the inverse square of the Gaussian beam size parameter $w(z)$ \cite{Vinet09}. 
% state linear spectral density or PSD here
However, as the beam spots are made larger, a larger fraction of the light power is lost over the edge of the reflective surface. This fraction is known as the \emph{clipping loss}. 
If one uses a beam with a more homogeneous radial intensity distribution than the commonly used LG$_{00}$ beam, the Brownian thermal noise level can be reduced without increasing the clipping loss. Such beams are often referred to as \emph{flat} beams. 
Candidate flat beams for gravitational wave detectors include mesa beams \cite{mesa}, optimised conical beams \cite{conical} and higher-order LG beams \cite{mours}. 

Higher-order LG modes have the advantage over other flat beams of being theoretically compatible with the spherical mirror surfaces that are currently used with the LG$_{00}$ mode. 
Equation \ref{eqn:lghxamp} shows the complex amplitude function for the helical higher-order mode set,
\begin{equation} 
u_{p,l}^{\textrm{hel}}(r,\phi,z) =\frac{1}{w(z)}\sqrt{\frac{2p!}{\pi(|l|+p)!}}\quad e^{
\left(2p+|l|+1\right)\Psi(z)}\\
\times \,\left(\frac{\sqrt{2}r}{w(z)}\right)^{|l|}L^{(|l|)}_{p}\left(\frac{2r^2}{w(z)^2}\right)
e^{-\textrm{i}k\frac{r^2}{2q(z)}+\textrm{i}l\phi},\\
\label{eqn:lghxamp}
\end{equation} 
where $p$ is the radial mode index, $l$ is the azimuthal mode index, 
$w$ is the beam radius, $k$ is the wave-number, $q$ is the complex
Gaussian beam parameter,
$L^{(|l|)}_{p}$ 
are the associated Laguerre polynomials, and 
$\Psi$ is the Gouy phase, given by 
\begin{equation}
\Psi(z)=\textrm{arctan}\left(\frac{\lambda z}{\pi w_0^2}\right).
\label{eqn:Gouyphase}
\end{equation}
The order of a higher-order LG mode is given by $2p+|l|$. 
An alternative sinusoidal form of LG modes with a sinusoidal amplitude dependence in
azimuthal angle can be used equally well.
In this article we will discuss the research performed by the Birmingham group towards evaluating the feasibility of LG mode technology for implementation in detectors of the advanced generation and beyond to reduce the test mass thermal noise limit. 

\section{Interferometric performance simulation study of LG modes}
Our initial step was to compare the interferometric performance of the LG$_{33}$ mode to the LG$_{00}$ mode in an optical layout similar to that proposed for the advanced detectors. 
By interferometric performance we mean the level of coupling of a number of variables to the measured phase (`phase noise analysis'), the ability to generate the required longitudinal and alignment control signals, and finally the maximum achievable detector sensitivity. 
The first consideration to ensure a fair analysis of the performance of the LG$_{33}$ mode was the beam sizes which should be compared. 
A realistic design for advanced detectors will push towards larger beam sizes on the test masses, irrespective of the spatial mode used, since the limiting Brownian noise level scales with the inverse square of the beam size. 
The limit on beam size can be enforced either by the point at which the clipping loss becomes unacceptably high, or the point at which cavities become unacceptably close to the unstable concentric design required for maximising beam sizes. 
For this study we considered the limit to be enforced by the clipping loss, which we fixed at 1\,ppm. 

The higher-order LG modes are more spatially extended, and thus have larger clipping losses than the LG$_{00}$ mode for a given beam spot size parameter.
%A different beam size must therefore be used for each different mode considered to give the fixed 1\,ppm clipping loss. 
The ratio between LG$_{00}$ spot size and LG$_{33}$ spot size required to give a 1\,ppm clipping loss on the same mirror is $\frac{w_{00}}{w_{33}}=1.64$. 
In the rest of this section we will compare the performance of three different configurations for a symmetric 3\,km cavity, to be referred to as the LG$_{33}$, LG$_{00}^{\textrm{\tiny{large}}}$ and LG$_{00}^{\textrm{\tiny{small}}}$ configurations. 
The LG$_{33}$ and LG$_{00}^{\textrm{\tiny{large}}}$ configurations both have a clipping loss of 1\,ppm at the cavity mirrors, thus the beam size at the mirrors for LG$_{00}^{\textrm{\tiny{large}}}$ configuration is factor of 1.64 larger than for the LG$_{33}$ configuration. 
%It is therefore necessary to use different cavity mirror curvatures for these two configurations. 
The  LG$_{00}^{\textrm{\tiny{small}}}$ configuration has the same spot size at the mirrors as the LG$_{33}$ configuration, but uses a LG$_{00}$ beam, and therefore has a lower clipping loss. 
The LG$_{00}^{\textrm{\tiny{small}}}$ configuration is less realistic for comparison with the LG$_{33}$, for the aforementioned reason that advanced detectors will be pushing towards larger beam sizes irrespective of spatial mode used, but we found it useful to include in the study to separate the effects due beam parameters and effects due directly to mode shape.

\begin{figure*}[htb]
\centerline{
\includegraphics[width=16cm,keepaspectratio]{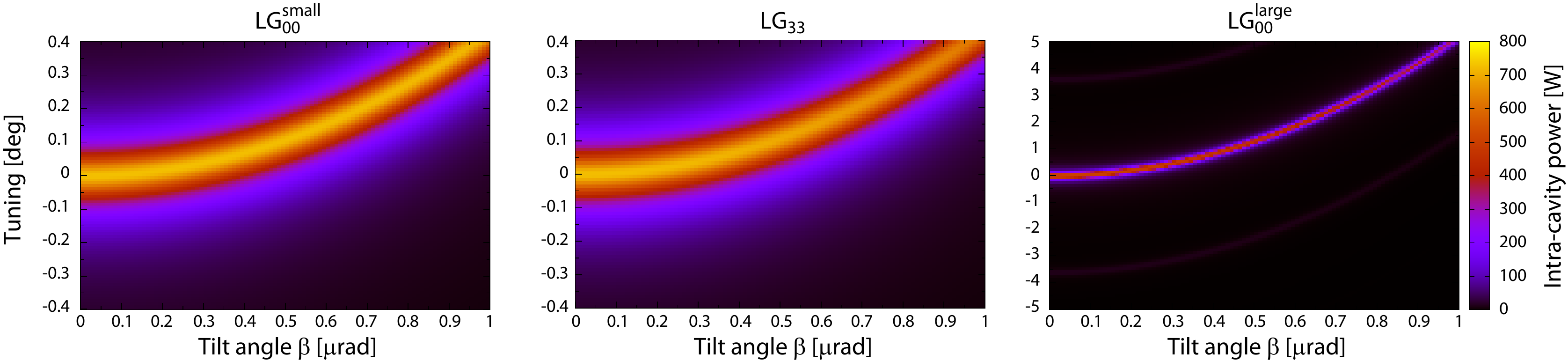}}
\caption{
Intra cavity power as a function of cavity end mirror tilt angle $\beta$ and longitudinal tuning $\phi$ 
for the three different configurations. 
The first two configurations show very similar results, but the LG$_{00}^{\textrm{\tiny{large}}}$ configuration shows 
a much stronger coupling from tilt to tuning (note the larger scale on the tuning axis in the LG$_{00}^{\textrm{\tiny{large}}}$ plot).
%Intra cavity power over tilt angle $\beta$ of the end
%mirror (EMX) and longitudinal tuning $\phi$ of a single cavity
%shown for all three configurations of interest. The first two
%configurations show almost the same coupling from tilt into
%longitudinal tuning which is more than an order of magnitude lower
%than the coupling of the third configuration.
}
\label{fig:tilttophase}
\end{figure*}

The first control signal investigated was the longitudinal error signal for a single cavity. 
For a Pound-Drever-Hall (PDH) modulation/demodulation error signal generation technique \cite{pdh}, the resulting longitudinal control signal was identical for all three of the considered configurations. 
The next step was an analysis of the coupling from end mirror tilt to longitudinal phase noise in a single cavity for each configuration. 
Figure \ref{fig:tilttophase} shows the results of the investigation for a 3\,km cavity based on the Advanced Virgo reference design in \cite{VIR-NOT-EGO-1390-330}. 
The LG$_{00}^{\textrm{\tiny{small}}}$ configuration performs slightly better than the LG$_{33}$ here, but of the two configurations with equal clipping losses, the LG$_{33}$ configuration performs significantly better than the LG$_{00}^{\textrm{\tiny{large}}}$ configuration. 
This is because the LG$_{00}^{\textrm{\tiny{large}}}$ configuration requires the cavity to be pushed closer to the unstable concentric design in order to achieve the necessary beam sizes at the mirrors.

\begin{figure*}[t]
\centerline{
\includegraphics[width=12cm,keepaspectratio]{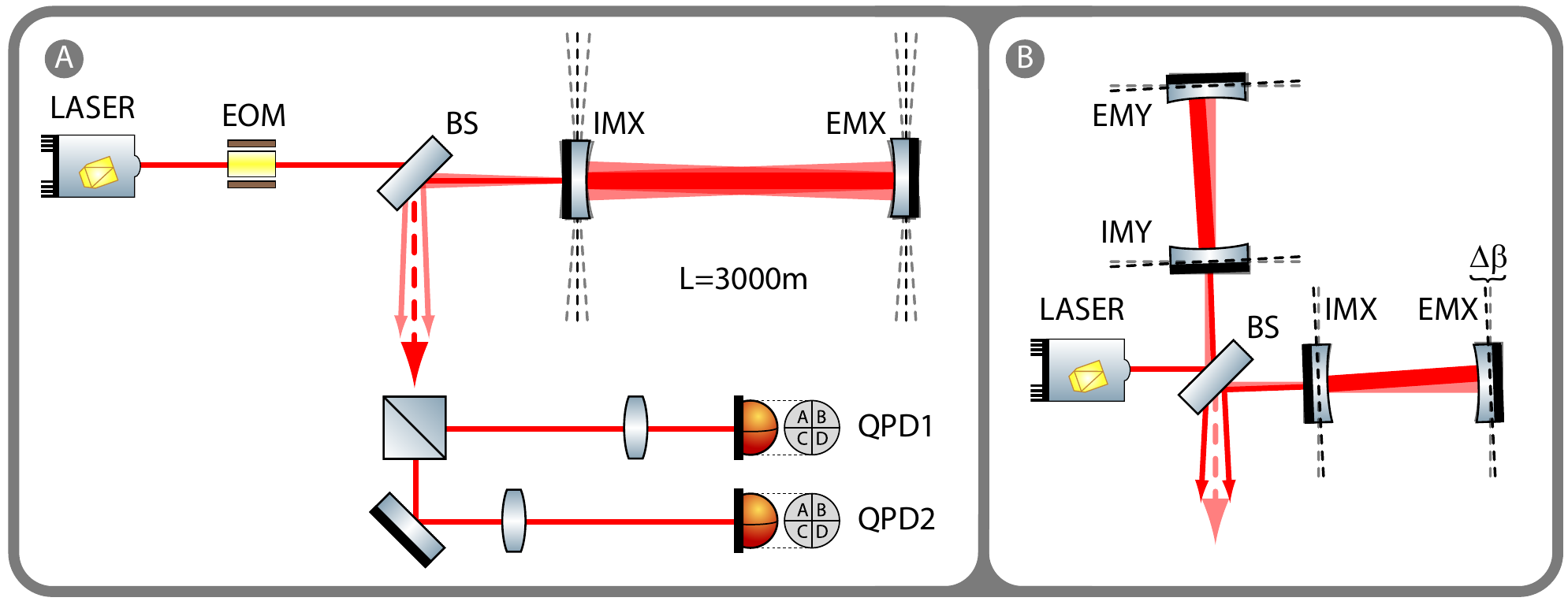}
} \caption{Two optical layouts used in the alignment analysis simulations. \textcircled{\scriptsize{\textbf{A}}} shows the single arm cavity alignment control scheme investigated, and \textcircled{\scriptsize{\textbf{B}}} shows the differential misalignment of arm cavities, for which coupling into dark port power was analysed.
%Different optical layouts used in the alignment
%analysis described in Section~\ref{sec:alignment_analysis}. Left
%\textcircled{\scriptsize{\textbf{A}}}: The generation of alignment
%error signals for a single arm cavity. Right
%\textcircled{\scriptsize{\textbf{B}}}: Michelson interferometer
%with differentially misaligned arm cavities to study the power
%coupling into the output port of the interferometer.
}
\label{fig:alignment}
\end{figure*}

As well as length sensing and control, angular sensing and control is crucial to maintaining stable operation and maximum sensitivity of a gravitational wave interferometer. 
To investigate how the LG$_{33}$ mode performs in this respect, an alignment scheme based on the Ward technique described in \cite{Ward} was designed for the same 3\,km cavities previously described. 
The optical layout for this scheme is shown in panel \textcircled{\scriptsize{\textbf{A}}} of figure \ref{fig:alignment}. 
Radio frequency sidebands are added to the beam by the electro-optic modulator (EOM), and the beam is passed into the cavity via a beam splitter. 
The reflected beam from the cavity is split in two, and each resulting beam is passed through a telescope and detected with a quadrant photodetector. 
These telescopes are designed such that the Gouy phase (see equation \ref{eqn:Gouyphase}) difference between the beams at each quadrant photodetector is 90$^\circ$, in order to provide the maximum possible orthogonality between alignment signals from the end mirror and the input mirror. 
The signal from each photodetector is demodulated by mixing with the initial radio frequency, with a demodulation phase chosen such as to maximise the slope of the error signal corresponding to the mirror for which the photodetector is required to sense the alignment. 
In order to match the conditions under which the alignment control system will be developed in practice as closely as possible, we tuned the parameters rather than using the theoretical optimum parameters. 

The alignment sensing figure of merit for each configuration can be summarized by its control matrix \cite{Mantovani08}. 
The elements of these control matrices are the slopes of the error signal at the working point as measured at quadrant photodetectors QPD1 and QPD2, for misalignments of the cavity input mirror IMX, and end mirror EMX, as shown in equation \ref{eqn:config}.
\begin{equation}
C_\mathrm{configuration}= \left(
\begin{array}{cc}
\sigma^\mathrm{IMX}_\mathrm{QPD1} & \sigma^\mathrm{EMX}_\mathrm{QPD1} \\
\sigma^\mathrm{IMX}_\mathrm{QPD2} & \sigma^\mathrm{EMX}_\mathrm{QPD2} \\
\end{array}
\right)\,\\
\label{eqn:config}
\end{equation}
%\hspace{-2cm}
The resulting control matrices for each of the three cavity configurations were as follows: 
\begin{eqnarray}
%\hspace{-1cm} 
%C_\mathrm{configuration}= \left(
%\begin{array}{cc}
%\sigma^\mathrm{IMX}_\mathrm{QPD1} & \sigma^\mathrm{EMX}_\mathrm{QPD1} \\
%\sigma^\mathrm{IMX}_\mathrm{QPD2} & \sigma^\mathrm{EMX}_\mathrm{QPD2} \\
%\end{array}
%\right)\,\\
%\hspace{-1cm} 
C_\mathrm{LG_{00}^{small}}=\left(
\begin{array}{cc}
  5.6152 & 0.0477 \\
  2.1607 & 3.5878 \\
\end{array}
\right)=\,5.6152
\left(%
\begin{array}{cc}
  1 & 0.009 \\
  0.385 & 0.639 \\
\end{array}
\right)\\
\hspace{-1cm} C_\mathrm{LG_{33}}=\left(
\begin{array}{cc}
  7.444 & 0.022 \\
  2.741 & 4.771 \\
\end{array}%
\right)=\,7.444
\left(%
\begin{array}{cc}
  1 & 0.003 \\
  0.368 & 0.641 \\
\end{array}
\right)\\
\hspace{-1cm} C_\mathrm{LG_{00}^{large}}=\left(
\begin{array}{cc}
  17.774 & 15.330 \\
  11.472 & 2.725 \\
\end{array}%
\right)=\,17.774
\left(%
\begin{array}{cc}
  1 & 0.862 \\
  0.645 & 0.153 \\
\end{array}
\right).
\end{eqnarray}
An ideal control matrix would be diagonal, since the off diagonal elements correspond to the presence of information from the unwanted mirror in a given photodetector signal. In this respect, the LG$_{33}$ and LG$_{00}^{\textrm{\tiny{small}}}$ configurations give control matrices which are much closer to the ideal than the  LG$_{00}^{\textrm{\tiny{large}}}$ configuration; the information from the two mirrors is better separated in these two configurations. 

\begin{figure}[t]
\centerline{
\includegraphics[width=9cm,keepaspectratio]{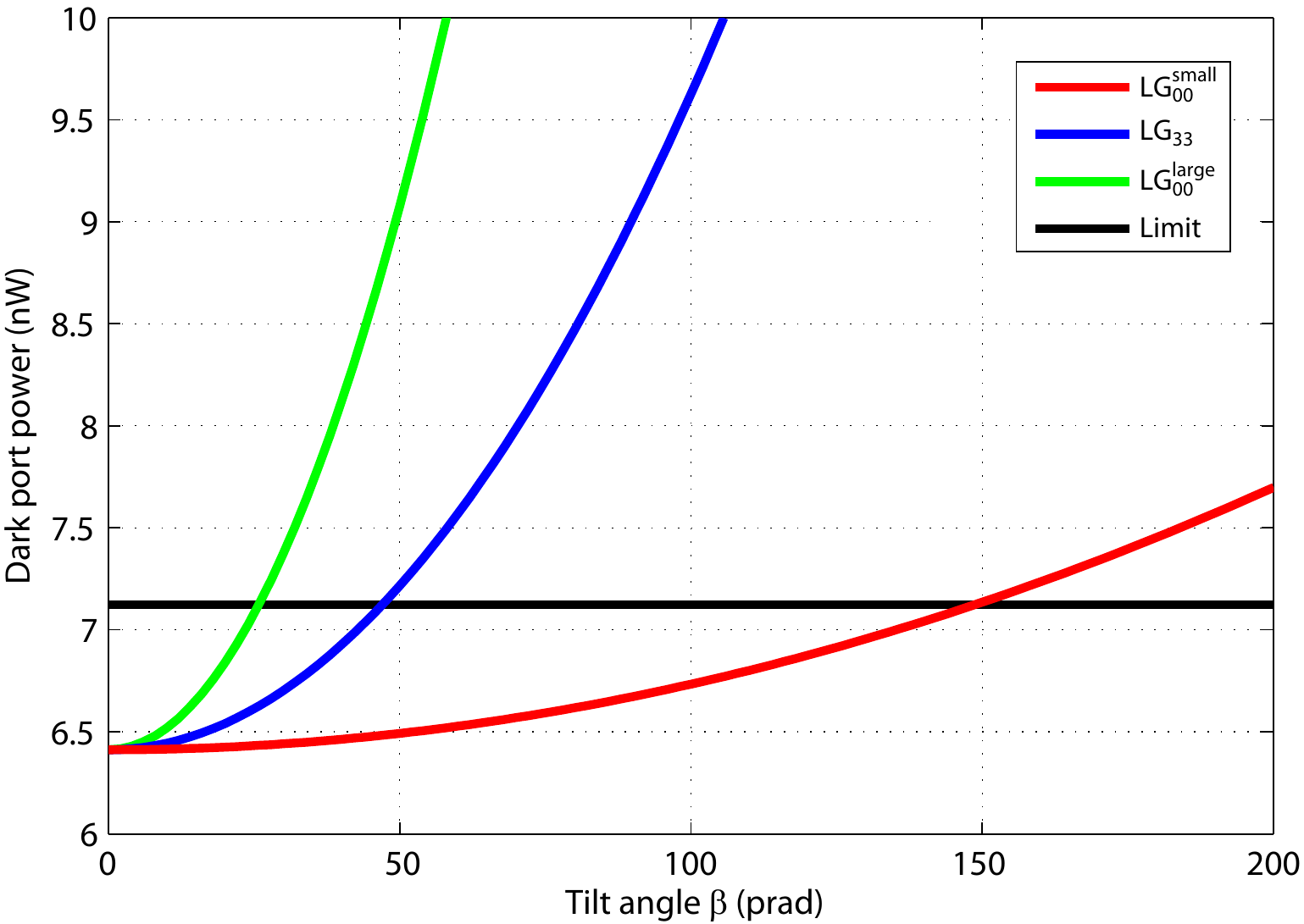}}
\caption{
Dark port power as a function on differential arm cavity misalignment for the three different configurations. 
The black line shows the limit calculated for dark port power based on a differential arm length requirement of 10$^{-15}$\,m 
for a dark fringe offset of 10$^{-12}$\,m. 
}
\label{fig:diffmisalign}
\end{figure}

The next aspect of the interferometric performance of the LG$_{33}$ mode to be investigated was the coupling of differential arm cavity misalignment to power at the interferometer output port. 
If the two arm cavities are differentially misaligned, the overlap of the two beams at the central beam splitter will not be complete, leading to a change in the light power present at the output port of the interferometer, as illustrated in panel \textcircled{\scriptsize{\textbf{B}}} of figure \ref{fig:alignment}. 
If this differential misalignment varies with time, the power measured at the dark port will also vary with time, producing a signal at the photodetector that is indistinguishable from gravitational wave signals. 
We obtained values for the output power enhancement due to differential misalignment for each of the three previously described cavity configurations, as shown in figure \ref{fig:diffmisalign}.
Comparing the simulation results with the reference limit (around 7$\times$10$^{-9}$\,W) calculated from a differential arm length requirement of 10$^{-15}$\,m \cite{Adhikari08} and a dark fringe offset of 10$^{-12}$\,m \cite{Abbott08}, we see that the LG$_{00}^{\textrm{\tiny{small}}}$ configuration performs best, but crucially that the LG$_{33}$ configuration also outperforms the LG$_{00}^{\textrm{\tiny{large}}}$ configuration. 
In summary, for all the interferometric performance aspects analysed, the LG$_{33}$ configuration performs significantly better than the LG$_{00}^{\textrm{\tiny{large}}}$ configuration. 
The  LG$_{00}^{\textrm{\tiny{small}}}$ performs as well or better than the LG$_{33}$ configuration, but it should be recalled that this will have a significantly higher thermal noise level, and for this reason is not representative of a likely configuration for advanced detectors. 
%Comparison of the dark port power when the arm cavities
%are differentially misaligned 
%(see
%Fig.~\ref{paper-cavity-alignment.eps}\,\textcircled{\scriptsize{\textbf{B}}})
%in reference to the dark fringe power that results from a
%differential arm length deviation of $10^{-15}$\,m while the
%interferometers dark fringe offset is $10^{-12}$\,m.

%Using the Advanced LIGO limit for differential arm length deviations of 10$^{-15}$\,m, and the planned dark fringe offset of 10$^{12}$\,m, we calculated a limit on the power increase at the output port of 7.124$\times$10$^{-9}$\,W. 

The final comparison between the performance of the LG$_{00}$ and LG$_{33}$ modes was in terms of the overall detector sensitivity. 
Several scenarios for using the LG$_{33}$ mode in Advanced Virgo detector were evaluated in \cite{Chelkowski09}, but here we just discuss a scenario which compares the sensitivity of the detector with LG$_{00}$ and LG$_{33}$ modes where both beams experience a 1\,ppm clipping loss at the cavity mirrors. 
The cavity mirror curvatures $RoC$ and beam sizes at the mirrors $w$ for the two configurations in this scenario are shown in table \ref{tab:sensitivity}.
In each case the detector sensitivity was calculated using a version of  Gravitational Wave Interferometer Noise Calculator (GWINC) \cite{gwinc} especially adapted for Advanced Virgo. 
Table \ref{tab:sensitivity} shows the results of the calculation for the two cases NS/NS, and BH/BH where the signal recycling detuning (SR det.)\cite{Hild07} was optimized for detection of signals from binary neutron star inspirals and binary black hole inspirals respectively. 
The figures of merit chosen were the effective detection ranges for the two signal sources, $\Gamma_\mathrm{NS/NS}$ and $\Gamma_\mathrm{BH/BH}$.
According to these results, the LG$_{33}$ mode provides a relative improvement of the inspiral ranges by around 20\,\%  and 25\,\% for signal recycling detunings of 750\,Hz and 300\,Hz respectively, compared to the LG$_{00}$ mode. 
This corresponds to a potential increase by up to a factor of 2 in the observable event rate of the Advanced Virgo detector by using the LG$_{33}$ mode instead of the LG$_{00}$ mode. The method and results of the investigation described in this section are presented in more detail in \cite{Chelkowski09}.

\begin{table*}[htbp]
    \begin{center}
%    \begin{ruledtabular}
        \begin{tabular}{|c||cccccc|}
\hline
Laser& SR det. & $w$  & $l_\mathrm{clip}$  & RoC & $\Gamma_\mathrm{NS/NS}$  & $\Gamma_\mathrm{BH/BH}$\\
mode & [Hz] & [cm] & [ppm] & [m] & [Mpc] & [Mpc]\\
\hline \hline
LG$_{00}$ & 750 & 6.47 & 1 & 1522.8 & 125.54 & 900.29\\
\hline
LG$_{33}$  & 750 & 3.94 & 1 & 1708.4 & 148.30 & 1142.6\\
\hline
LG$_{00}$ & 300 & 6.47 & 1 & 1522.8 & 130.08 & 580.41\\
\hline
LG$_{33}$  & 300 & 3.94 & 1 & 1708.4 & 163.07 & 714.74\\
\hline
        \end{tabular}
\caption{Relevant parameters and results of the GWINC simulation
analysis of an Advanced Virgo design which considers identical clipping loss $l_\mathrm{clip}$ of 1\,ppm at
the arm cavity mirrors for both the LG$_{00}$ and LG$_{33}$ modes.}
        \label{tab:sensitivity}
 %       \end{ruledtabular}
    \end{center}
\end{table*}

\section{Table-top demonstrations}
Following the positive results of the simulation investigation into the interferometric performance of the LG$_{33}$ mode, we proceeded to perform table-top experiments to verify these results. 
The first step was to generate LG modes with a reasonably high purity for use in the experiments. 
This was achieved using a Holoeye liquid-crystal on Silicon spatial light modulator (SLM), as for example in \cite{matsumoto}. 
Upon reflection from the surface of the SLM, a beam can be imprinted with a transverse phase pattern determined by the user. 
By choosing the phase pattern to be that of a LG$_{33}$ mode, i.e. three azimuthal phase vortices and three radial phase jumps of $\pi$ for the helical case, one can achieve a partial conversion from the LG$_{00}$ mode to the LG$_{33}$ mode. 
%\begin{figure}[htb]
%\begin{center}
%\includegraphics[width=0.48\textwidth,keepaspectratio]{LGmode-lin-cav.pdf}
%\end{center}
%\caption{The experimental setup for mode cleaning a SLM generated higher-order LG beam. 
%The \tem input beam is converted to a higher-order LG beam
%by the SLM. The resulting beam is passed through a mode-matching telescope into the linear cavity. 
%The transmitted light is used to generate an error signal which is fed back to the PZT attached to the 
%curved end mirror to control the length of the cavity. The transmitted beam is simultaneously imaged 
%on the CCD camera.}
%\label{fig:tabletopexp}
%\end{figure}
\begin{figure}[h]
\begin{minipage}{20pc}
\includegraphics[width=20pc]{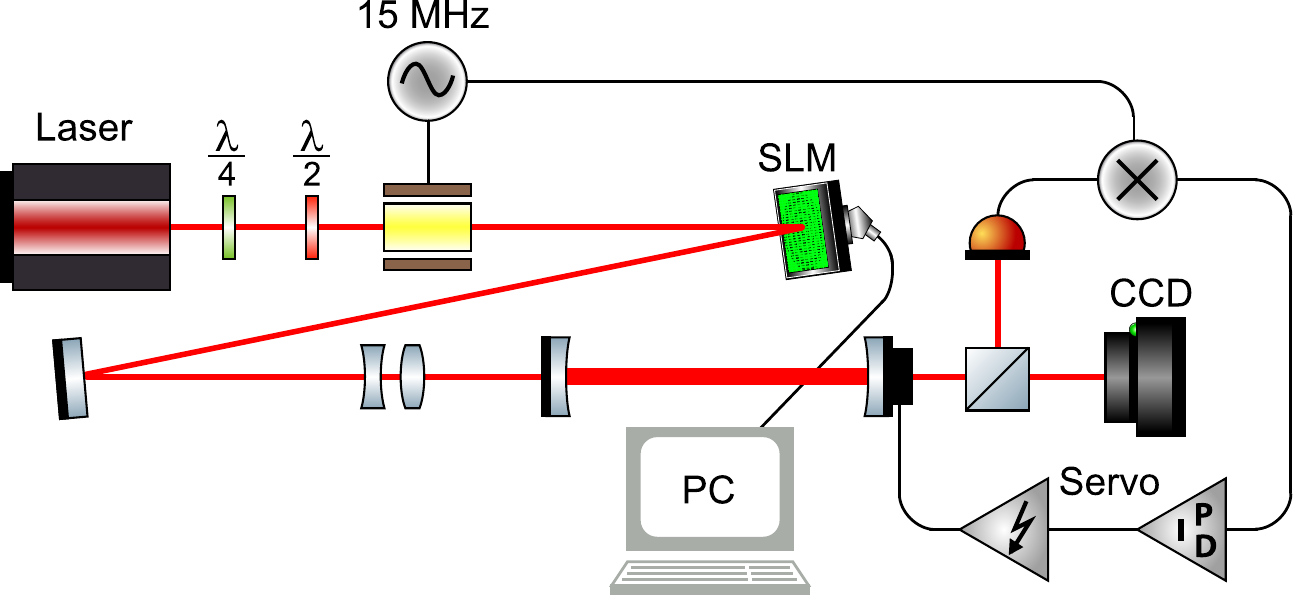}
\caption{\label{fig:tabletopexp}The experimental setup for mode cleaning a SLM generated higher-order LG beam. 
The input beam is converted to a higher-order LG beam
by the SLM, then passed via a mode-matching telescope into the linear mode cleaner cavity. 
The transmitted light is used to generate an error signal which is fed back to the PZT attached to the 
curved end mirror to control the length of the cavity, and simultaneously imaged 
on the CCD camera.}
\end{minipage}\hspace{2pc}%
\begin{minipage}{16pc}
\includegraphics[width=16pc]{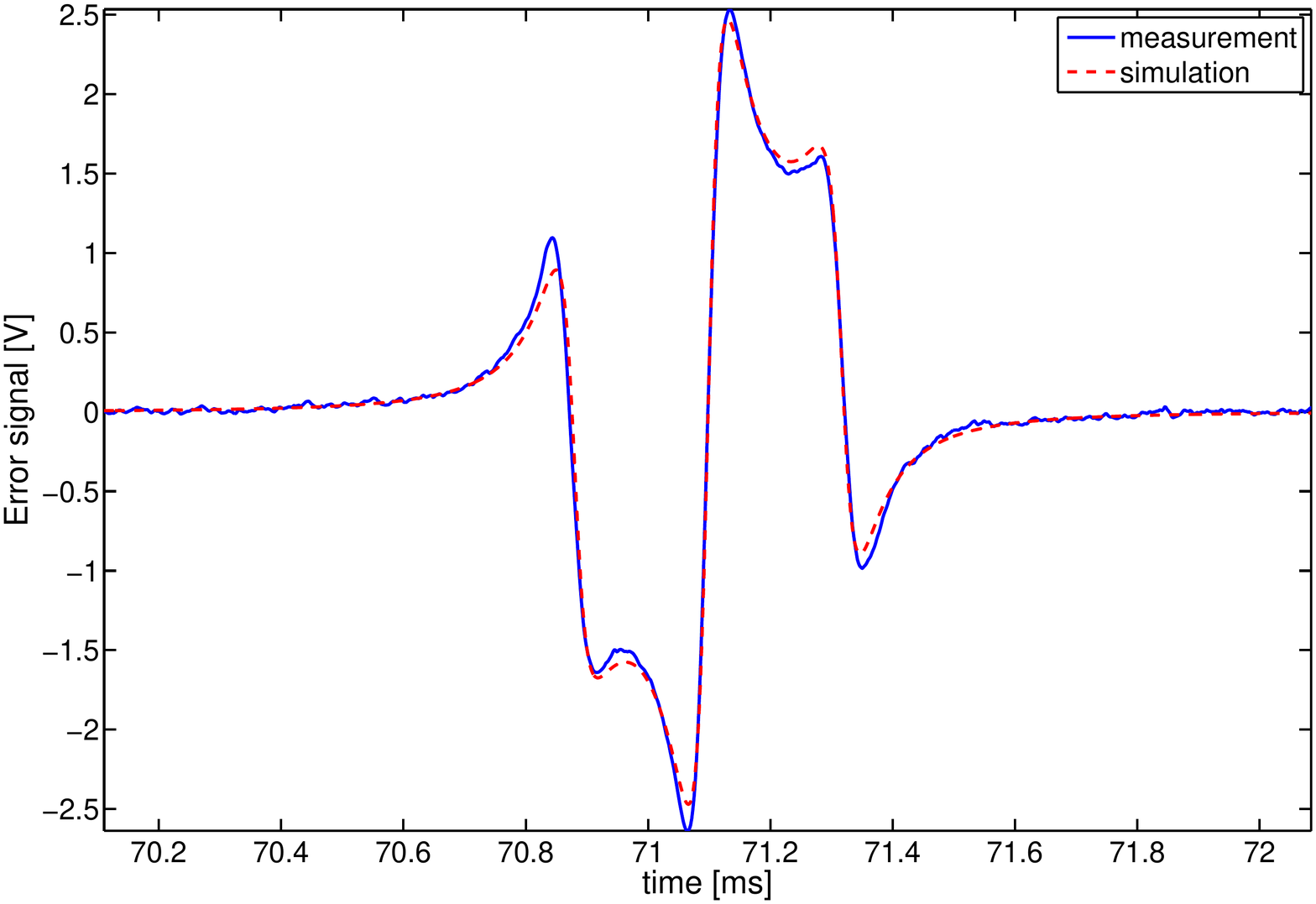}
\caption{\label{fig:pdh}The blue trace and the red dashed trace show the measured sinusoidal LG$_{33}$ beam and simulated LG$_{00}$ PDH error signal respectively, for the linear mode cleaner cavity set up as shown in figure \ref{fig:tabletopexp}. 
The primary 
features are identical, as predicted in \cite{Chelkowski09}.}
\end{minipage} 
\end{figure}

The experimental setup shown in figure \ref{fig:tabletopexp} was then used to investigate the length control signals that could be generated for the resulting beam. 
The 1064\,nm laser light was passed through an EOM to add 15\,MHz sidebands to the light, for generating the PDH error signal for length control of the mode cleaner cavity. 
The light was then partially converted to the LG$_{33}$ mode by interaction with the SLM, and passed into the mode cleaner cavity via a mode matching telescope. The light transmitted through the mode cleaner cavity was simultaneously measured with a photodiode and a CCD camera. 
The photodiode signal was mixed down with the 15\,MHz signal to give the required error signal, which was then fed back to a Piezo-electric transducer attached to one of the mode cleaner mirrors, via two amplifier stages. 
In this way the length of the cavity was controlled to keep it in the resonant condition for any given mode order. The CCD measurement of the transmitted beam was then used for analysing the mode purity. 
The mode cleaner is a 2 mirror flat-concave cavity of length 21\,cm and with a measured finesse of 172. 
%
%\begin{figure}[htb]
%\begin{center}
%\includegraphics[width=0.38\textwidth,keepaspectratio]{sinusoidalPDH.pdf}
%\caption{The blue trace shows the PDH error signal from the linear cavity, set up as shown in figure 
%\ref{fig:LGmode-lin-cav}, with a sinusoidal \lag input beam. The red dashed trace shows the PDH 
%error signal for the same optical setup as simulated in the frequency domain simulation software 
%Finesse \cite{Finesse}. While there are small discrepancies between the two traces, the primary 
%features are identical, as predicted in \cite{Chelkowski09}.}
%\label{fig:pdh}
%\end{center}
%\end{figure}

We verified the first result of the initial simulations by recording the PDH error signal generated using the LG$_{33}$ mode, shown in figure \ref{fig:pdh}. 
This signal was equivalent for the LG$_{33}$ and LG$_{00}$ modes as predicted, and therefore allowed us to control the cavity on resonance for the order 9 LG$_{33}$ mode. 
Demonstrating robust and repeatable feedback control of an optical cavity was a significant step in validating the higher-order LG mode operation of interferometers.

\begin{figure}[htb] 
\begin{center} 
\includegraphics[width=0.21\textwidth,keepaspectratio]{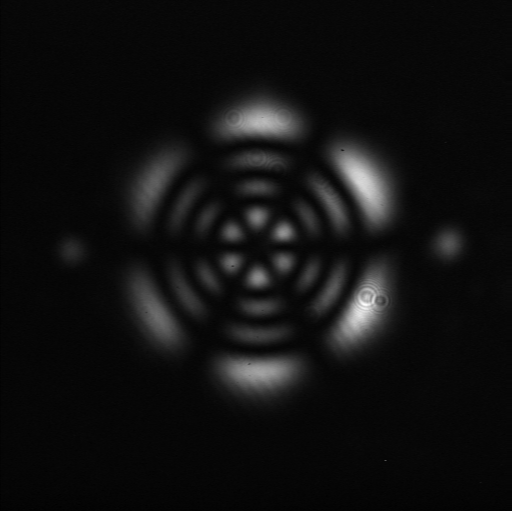}\hspace{0.05cm}
\includegraphics[width=0.21\textwidth,keepaspectratio]{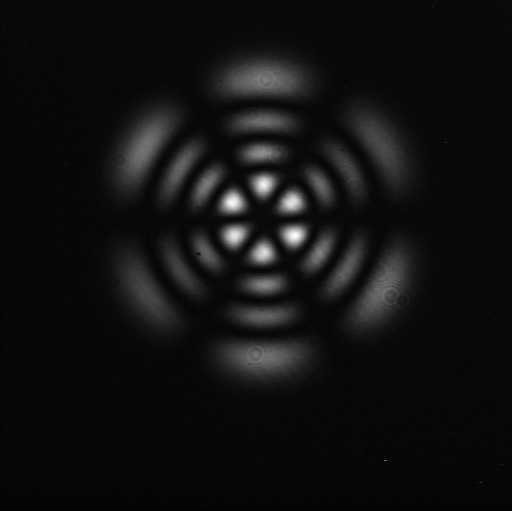}\vspace{0.1cm}
\hspace{1cm}
\includegraphics[width=0.21\textwidth,keepaspectratio]{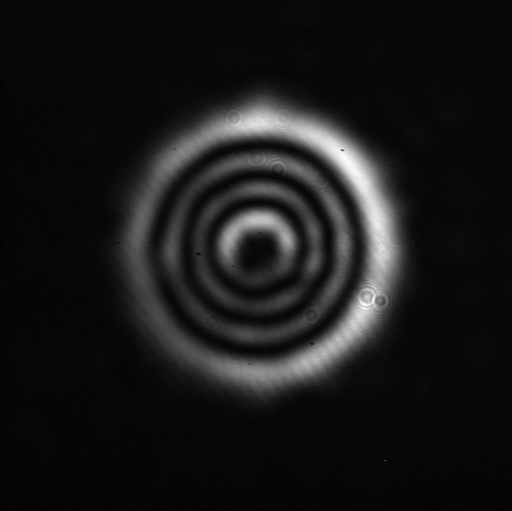}\hspace{0.05cm}
\includegraphics[width=0.21\textwidth,keepaspectratio]{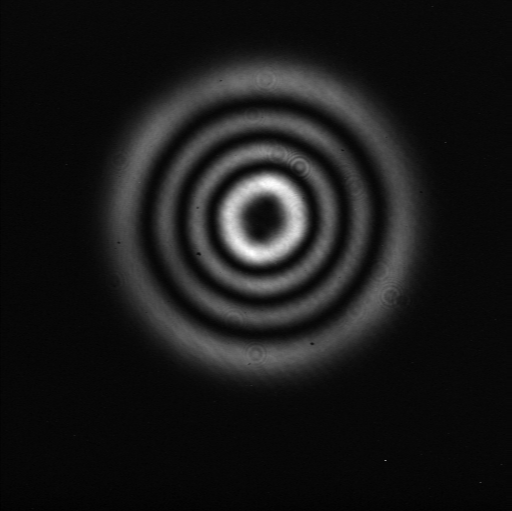}\vspace{0.1cm}
\caption{The measured intensity patterns of the sinusoidal (left pair) and helical (right pair)
 LG$_{33}$ beams before (left of pairs) and after (right of pairs) transmission through
the linear mode cleaner. 
%The increase in mode purity upon transmission is already
%evident in the increased symmetry. The remaining asymmetry apparently is a result of the inaccuracy in 
%the manual alignment of the input beam to the mode cleaner. This effect is the same for both images
%but more visually apparent in the case of the helical mode.
} 
\label{fig:modepics}
\end{center} 
\end{figure} 

Figure \ref{fig:modepics} shows the beams before and after transmission through the mode cleaner cavity. 
It is immediately apparent that the transmitted beams exhibit greater circular symmetry than the input beams, as would be expected due to the filtering out of other spatial modes. 
We used the CCD measurement of the transmitted beam intensity pattern to estimate its mode content. 
This required fitting the beam size parameter to calibrate the CCD, and subtracting an ideal LG$_{33}$ mode intensity pattern, with beam size given by the result of the fit, from the measured data. 
These residuals were then analysed by comparing the measured data with a numerical simulation of the setup which included some misalignment into the cavity, performed using the interferometer simulation Finesse \cite{Finesse}.
For a Finesse model of the mode cleaner setup which included an input beam misalignment of -100\,$\mu$rad in horizontal and 60\,$\mu$rad in vertical, the residual between the transmitted mode and an ideal LG$_{33}$ matched the measured data extremely well. 
%The residuals between the ideal LG$_{33}$ mode and the measured cavity input beam, measured transmitted beam, and simulated transmitted beam respectively are shown for the sinusoidal mode case in figure \ref{fig:residuals}.
By performing a modal decomposition of the transmitted field in the simulation, we estimated the mode content of the transmitted beam, as shown in table \ref{tab:lg33decompose}. 
\begin{table}[htb]
\begin{center}
\begin{tabular}{|l|c|c|c|c|c|c|}
\hline
$u^{\rm sin}_{l p}$ mode & 3, 3 & 4, -1 & 2, -5& 4, 1& 2, 5& other\\
\hline
power &  99\% & 0.4\% & 0.3\% & 0.1\% & 0.1\% &  $<10$\,ppm \\
\hline
\end{tabular}
\caption{Results of a modal decomposition of the transmitted sinusoidal LG$_{33}$ beam in the numerical model of the linear mode cleaner, with an input beam misalignment of -100\,$\mu$rad in the 
horizontal axis, and 60\,$\mu$rad in the vertical axis. 
Most of the beam is in the sinusoidal LG$_{33}$ mode, with the rest almost entirely found 
in other modes of order 9.}\label{tab:lg33decompose}
\end{center}
\end{table}
The analysis showed that the transmitted beam was to 99\,\% in the LG$_{33}$ mode for both the helical and sinusoidal cases. 
%Since the light transmitted through the cavity can be considered to be a sum of eigenmodes of the cavity, the simulated setup
With an estimate of the transmitted mode purity, together with measurements of the cavity finesse and throughput efficiency, we were also able to estimate the LG$_{33}$ mode content of the input beam to be 66\,\% for the helical beam, and 51\,\% for the sinusoidal beam. A detailed description of the method and results of this experiment can be found in \cite{Fulda10}.
%\begin{figure}[htb]
%\begin{center}
%\includegraphics[scale=0.315,viewport=162 0 530 300,clip]{lg33sin-lmc-in-residual-pp.png}
%\includegraphics[scale=0.315,viewport=163 0 530 300,clip]{lg33sin-lmc-out-residual-pp.png}
%\includegraphics[scale=0.315,viewport=165 0 530 300,clip]{lg33_align_residual-pp.png}
%\caption{Residuals from best fits between intensity patterns and a theoretically ideal sinusoidal 
%LG$_{33}$ intensity pattern. From left to right: the residual for the measured input LG$_{33}$ beam, the residual for the measured 
%output LG$_{33}$ beam, the residual for an output LG$_{33}$ pattern generated with a numerical model including a
%misalignment of the input beam to the cavity.}
%\label{fig:residuals}
%\end{center}
%\end{figure}

%Access to the phase and amplitude of the beam would be necessary to make a direct measurement of mode purity, however this was not possible in our case using just a CCD camera.

\section{LG mode degeneracy investigation}
One important difference between higher-order LG modes and the fundamental LG$_{00}$ mode is that only the LG$_{00}$ mode is unique in its mode order. 
For each higher-order mode there exists at least one other mode of the same order. 
Since the mode filtering effect of optical cavities relies principally on the round trip Gouy phase difference between different mode orders, they cannot therefore filter out all other modes than the LG$_{33}$ when on resonance for order 9. 
For this reason we call the other order 9 modes degenerate with the LG$_{33}$ mode. 

This degeneracy of higher-order modes has serious implications for their application in gravitational wave interferometers.  
In the initial simulations reported in \cite{Chelkowski09} the interferometer mirrors were modelled as perfect spherical curved mirrors. 
In reality however, the mirrors will have some small deviations from  perfect spherical surfaces which can cause coupling from the LG$_{33}$ mode into other modes of order 9. 
If we consider this process occurring within the arm cavities of a gravitational wave interferometer, the degeneracy of the modes of order 9 will mean that the mode content of the circulating beam may have a significantly reduced proportion of LG$_{33}$, the remainder being made up principally of other order 9 modes. 
If the mode content of both arm cavities is different (which is likely to be the case if the coupling between modes is driven by the randomly oriented mirror surface distortions) the modal overlap at the beam splitter will be imperfect and the output port contrast will be reduced.

The method and results of a detailed numerical and analytical investigation into the effect of mirror surface distortions on the purity of LG modes within optical cavities were presented at the 2011 Amaldi meeting \cite{Bondproceedings}, and in \cite{Bond11}, and another study on this topic is presented in \cite{Hong11}. 
One of the most important outcomes from the work described in \cite{Bond11} was the derivation of an analytical formula for predicting the amount of coupling between different LG modes upon reflection from a mirror, based on the spatial features of mirror surfaces as described by Zernike polynomial functions. 
In the limit that the height of the surface distortions is much smaller than 
the wavelength of the light, coupling between an incident mode LG$_{pl}$ and a reflected mode LG$_{p'l'}$ is only significantly caused by Zernike polynomials Z$_n^m$ which satisfy the condition $m=|l-l'|$. 
Based on this result, it was possible to propose limits on the heights of the most important lower-order Zernike polynomial present on mirror surfaces, in order to achieve a circulating mode purity of over 99.9\,\% \cite{Bond11}.
%Since the injected mode was a pure LG$_{33}$ mode, and there was no coupling from the LG$_{33}$ mode into other modes of the same order from mirror surface distortions, the initial simulations did not consider this effect.

\section{Ongoing experimental investigations}
Following the simulation studies and table-top experiments, we continue to pursue experimental tests of LG mode technology. 
Currently, this involves evaluating the performance of the LG$_{33}$ mode in a suspended 10\,m cavity at the Glasgow prototype facility, and the development of a high-power LG$_{33}$ mode laser beam at the Hannover high-power laser facility.

Procession from table-top experiments to a prototype experiment with suspended mirrors is a standard feature of the development of new technologies for gravitational wave detectors, in order to ensure compatibility with realistic interferometer subsystems and requirements. 
Experiments at the Glasgow prototype also give us a way to investigate the LG mode degeneracy problem heretofore only looked at theoretically and numerically. 
Though we did not observe the detrimental effects of the mode degeneracy problem in our table-top experiment we believe this is most likely due to the relatively small beam sizes used on the mirrors, and the low cavity finesse in comparison to the advanced detector arm cavities. 
However, it is clear that LG modes must also work with larger beam sizes in order to provide the thermal noise benefits that make them attractive in the first place. 
LG modes must also be compatible with cavities of similar finesse to the advanced detector arm cavities. 
We have therefore been setting up an experiment at the Glasgow 10\,m prototype facility which uses larger beam sizes in a higher finesse cavity. 

 \begin{figure}[h]
%\begin{minipage}{12pc}
\begin{center}
\includegraphics[width=17pc]{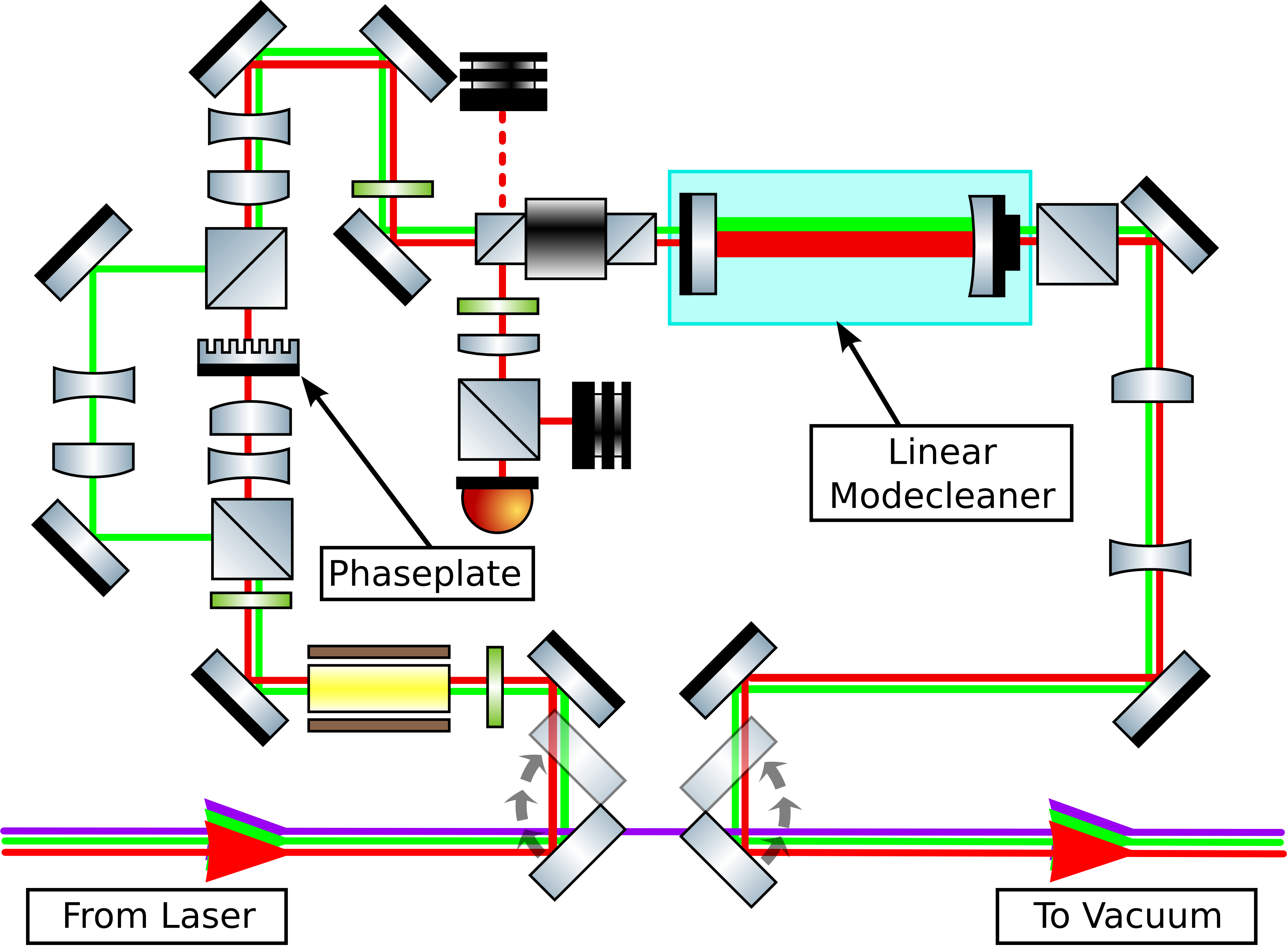}
\hspace{0.5cm}
%\caption{\label{fig:jifplan}caption.}
%\end{minipage}\hspace{6pc}%
%\begin{minipage}{20pc}
\includegraphics[width=17pc]{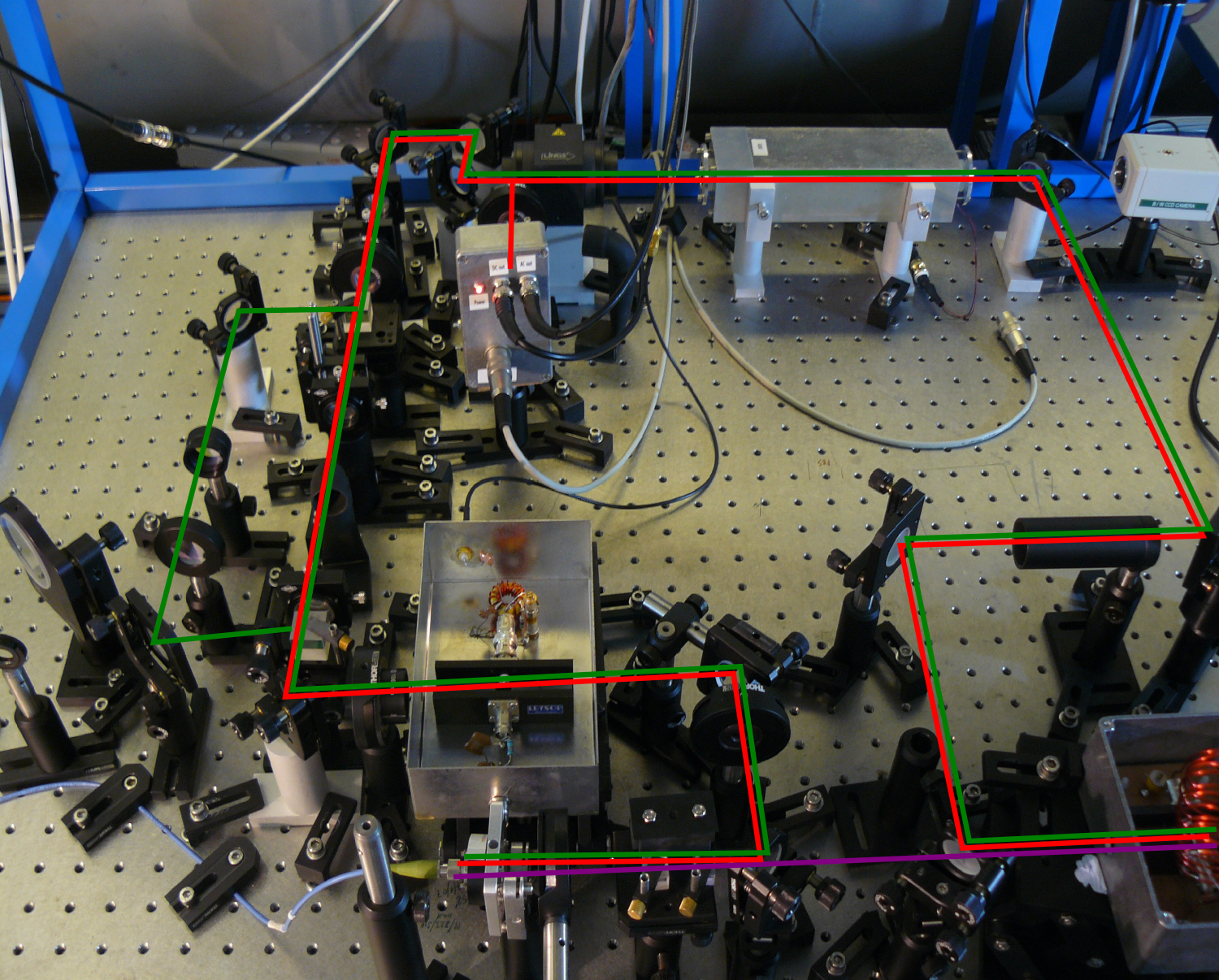}
\caption{\label{fig:jiftable}The laser mode conversion path on the Glasgow 10\,m prototype input optics bench. The red line shows the LG$_{33}$ mode path, and the green line shows the LG$_{00}$ path that bypasses the phaseplate. The purple line shows the original LG$_{00}$ laser path.}
%\end{minipage} 
\end{center}
\end{figure}
In this experiment we will use an etched diffractive optic to facilitate the conversion from LG$_{00}$ to LG$_{33}$, similar to that used in \cite{DOE}, rather than the SLM used in the table-top experiments. 
Though the SLM was very useful for prototyping purposes to investigate the effects of different phase profiles, a fixed transmissive optic allows to generate a higher mode purity with more stability than the SLM. 
The setup for the LG$_{33}$ mode conversion bench is shown in figure \ref{fig:jiftable}. 
The beam is picked off from the previously used laser path (shown in purple) after a fibre output coupler, and directed with a flip mirror into the LG mode conversion area. 
The red path shows the LG$_{33}$ mode conversion path, in which the beam passes through an EOM for generating the control sidebands, and the etched diffractive phaseplate. 
The resulting beam is then transmitted through a linear mode cleaner, which is feedback controlled using the PDH method to remain on resonance for mode order 9. 
This serves to increase the mode purity of the beam, as demonstrated in \cite{Fulda10}. 
The transmitted beam is then passed back into the previously used laser path using another flip mirror, and on towards the suspended cavity.
% where it subsequently passes another EOM for generating control signals for the suspended cavity, a Faraday isolator, and numerous lenses for matching the beam parameters to the suspended cavity eigenmode. 
The green path is a phaseplate bypass path, which enables us to alternatively operate the linear mode cleaner with the LG$_{00}$ mode. 
%this is essential to ensure a fair comparison between the performance of the LG$_{00}$  and LG$_{33}$ modes in the suspended cavity. 
Since the light transmitted by the mode cleaner must be defined by the cavity eigenmodes, the transmitted LG$_{00}$  and LG$_{33}$ beams should have the same alignment and mode matching relative to the suspended cavity. This enables us to make a valid assessment of the relative performance of the two modes within the suspended cavity. 

The experimental work on LG modes at the Glasgow prototype is still under way, though we will soon have results about the performance of the LG$_{33}$ mode within a cavity of finesse in the region of 600 - around the same finesse foreseen for the advanced detector arm cavities. 
We plan to analyse the length control signals for the cavity, as well as the transmitted beam intensity pattern, in order to ascertain the level of coupling between LG modes that occurs in the cavity. 
Analysis of the circulating mode purity, together with measured data of the cavity mirror surfaces, should enable us to verify the results of the theoretical investigation into the LG mode degeneracy problem described in \cite{Bond11}. 
With this information, we should be in a position to make well informed decisions on the suitability of LG mode technology for future detectors, based on the surface quality of the available optics. 

Another crucial requirement for LG mode technology to be useful in future detectors is the ability to generate high-power and high-purity LG mode laser beams for the interferometer input. 
It is planned to use laser powers in excess of 100\,W in the advanced detectors in order to bring the shot noise level down, so any prospective LG mode light source must also be able to reach comparable power levels. 
In collaboration with the members of the AEI in Hannover that developed the new 200\,W laser system for Advanced LIGO \cite{winkelmann11}, we are currently setting up an experiment to investigate the technical challenges of generating LG modes at such high powers. 
High-power beams from tens to hundreds of Watts will be passed through etched diffractive phaseplates. Following the example of our very first experiment on LG modes described in section 3, the LG beams that are generated will be then filtered by means of a linear mode cleaner, and their subsequent purity analysed.
This work is also still at a preliminary stage, but we expect to be able to report results in the near future.

\section{Summary}
We have investigated the suitability of higher-order LG mode technology for use in future gravitational wave interferometers for reducing the test mass thermal noise, from many different angles. 
An initial simulation study showed that the longitudinal and angular control performance of the LG$_{33}$ mode in an Advanced Virgo-like interferometer was significantly better than the LG$_{00}$ mode with the same clipping losses at the cavity mirrors. 
It was also shown that increases in the observable event rates of Advanced Virgo by up to a factor of 2 could be gained by using the LG$_{33}$ mode in place of the LG$_{00}$ mode. 
Table-top experiments verified the equivalence of the LG$_{33}$ mode and LG$_{00}$ mode for PDH longitudinal control signal generation in an optical cavity, and demonstrated the mode cleaning of SLM generated LG$_{33}$ beams to reach a mode purity of over 99\,\%. 
More detailed theoretical investigations into the effects of mirror surface distortions on LG mode purity within optical cavities led to the derivation of an analytical formula for the coupling between LG modes caused by specific mirror distortions. 
Using this formula, mirror specifications required for compatibility with the LG$_{33}$ mode were subsequently derived. 
We are currently engaged in a collaborative effort with the University of Glasgow to verify the results of these theoretical investigations, by means of experiment within the 10\,m prototype facility in Glasgow. 
We are also currently investigating the feasibility of generating high-power and high-mode purity LG$_{33}$ beams at the Hannover high-power laser facility.
Higher-order LG modes can potentially offer a significant improvement in the sensitivity of future gravitational wave interferometers: our research program will evaluate how much of that potential is likely to be realisable in practice.
\ack We thank B. Barr, A. Bell, S. Huttner, J. Macarthur, B. Sorazu and K. Strain for providing the opportunity and helping to perform the LG mode experiments at the Glasgow 10\,m prototype, and we thank C. Bogan, P. Kwee and B. Willke for the same assistance in the high-power LG modes experiment in Hannover. This document has been assigned the LIGO document number LIGO-P1100146. 
%\subsection{Using \BibTeX}
%We highly recommend the {\ttfamily\textbf\selectfont iopart-num} \BibTeX\ package by Mark~A~Caprio \cite{iopartnum}, which is included with this documentation.

%
%\begin{figure}[h]
%\begin{minipage}{14pc}
%\includegraphics[width=14pc]{name.eps}
%\caption{\label{label}Figure caption for first of two sided figures.}
%\end{minipage}\hspace{2pc}%
%\begin{minipage}{14pc}
%\includegraphics[width=14pc]{name.eps}
%\caption{\label{label}Figure caption for second of two sided figures.}
%\end{minipage} 
%\end{figure}
%
%\begin{figure}[h]
%\includegraphics[width=14pc]{name.eps}\hspace{2pc}%
%\begin{minipage}[b]{14pc}\caption{\label{label}Figure caption for a narrow figure where the caption is put at the side of the figure.}
%\end{minipage}
%\end{figure}
%
%Using the graphicx package figures can be included using code such as:
%\begin{verbatim}
%\begin{figure}
%\begin{center}
%\includegraphics{file.eps}
%\end{center}
%\caption{\label{label}Figure caption}
%\end{figure}
%\end{verbatim}

\end{document}